# FAST IR ARRAY DETECTOR FOR TRANSVERSE BEAM DIAGNOSTICS AT DAΦNE

A. Bocci
Centro National de Aceleradores, University of Seville, C/Thomas Alva Edison 7, 41092 Seville, Spain
M. Cestelli Guidi, A. Clozza, A. Drago, A. Grilli, A. Marcelli, A. Raco, R. Sorchetti
INFN – Laboratori Nazionali di Frascati, Via Enrico Fermi 40, I-00044 Frascati RM, Italy
L. Gambicorti
Istituto Nazionale di Ottica Applicata – CNR, Largo Fermi 6, 50125 Firenze, Italy
A. De Sio, E. Pace
Università degli Studi di Firenze, Largo Fermi 2 , 50125 Firenze, Italy
J. Piotrowski
VIGO System SA, 3 Swietlikow Str. 01-389 Warsaw, Poland

*Abstract*

At the *Laboratori Nazionali di Frascati* of the *National Institute of Nuclear Physics* (INFN) an infrared (IR) array detector with fast response time has been built and assembled in order to collect the IR image of $e^-/e^+$ sources of the DAΦNE collider. Such detector is made by 32 bilinear pixels with an individual size of 50x50 μm$^2$ and a response time of ~1 ns. In the framework of an experiment funded by the INFN V$^{th}$ Committee dedicated to beam diagnostics, the device with its electronic board has been tested and installed on the DAΦNE positron ring. A preliminary characterization of few pixels of the array and of the electronics has been carried out at the IR beamline SINBAD at DAΦNE. In particular the detection of the IR source of the e$^-$ beam has been observed using four pixels of the array acquiring signals simultaneously with a four channels scope at 1 GHz and at 10 Gsamples/s. The acquisition of four pixels allowed monitoring in real time differences in the bunch signals in the vertical direction. A preliminary analysis of data is presented and discussed. In particular we will outline the correlation between signals and displacements of the source occurring with bunch refilling during a complete shift of DAΦNE.

## INTRODUCTION

Diagnostic for particle accelerators is a fundamental tool of colliders and synchrotron radiation rings. In particular, the transverse diagnostic of beams is an important issue for low emittance machines. Source images collected by synchrotron light monitors are typically used to measure the vertical size of the beam and to calculate the vertical emittance. In particular fast devices that allow bunch-by-bunch and turn-by-turn transverse images of particle beams, although challenging are fundamental tools for ring diagnostics. Only recently a gateable intensified charge-coupled device (ICCD) camera has been used to get the image of the transverse profile for a single bunch of electrons and positrons in the PEP-II storage rings [1].

At present the availability of fast IR detectors in the mid-IR range with sub-nanosecond response times allows the real time detection of sources, such as electron or positron bunches in a collider [2]. In such a way it appears possible a real time monitor of the bunch-by-bunch transverse size of the source at IR wavelengths, with the possibility not only to detect source instabilities but also identify its position in the bunch train.

At DAΦNE, we used an array detector with a nanosecond response to detect the image of the $e^-/e^+$ sources at IR wavelengths. The device with its electronic board has been assembled for the installation on the positron ring at DAΦNE. Preliminary measurements of few pixels of the array and of electronics have been carried out at the IR SINBAD beamline of DAΦNE. In the first tests, the detection of the IR emission of the electron beam has been detected with the array acquiring simultaneously signals with a fast scope using only four pixels. Nevertheless, the acquisition of four pixels allowed monitoring in real time differences in bunch signals in the vertical direction. In this contribution the experimental set-up of measurements is outlined, measurements and experimental results are presented. Finally future opportunities of this new diagnostic method are discussed.

## EXPERIMENTAL SET-UP

Recently a new device has been designed and built for dedicated measurements of fast images of particle beams in collaboration with the VIGO System company. The device consists of a fast array detector with 2x32 pixels, each characterized by a size of ~50x50 μm$^2$ and a response time of ~1 ns. This device has been assembled with an interface board to attempt a first transverse diagnostics of the positron bunches of DAΦNE at IR wavelengths.

For these preliminary measurements we built an interface board with pixels of the array connected by gold bonding wires to the board and finally to the input of an analog electronic board. Because of the short distance between two consecutive pixels of the array, i.e., ~ 20-25 μm, for purely technical reasons, we were able bonding only two near non-consecutive pixels of the two lines of the array. Consequently, the pixel pitch between two consecutive pixels bonded was ~150 μm.

The dedicated analog electronic board with 64 channels and a bandwidth ≥1 GHz per channel has been designed to amplify signals with a gain of ~ 50-55 dB. This gain is function of the power supply voltage of monolithic amplifiers used to amplify the signal. Each input channel of the electronic allows also biasing photoconductive pixels of the detector with a constant bias current per pixel. For the first test we used SINBAD: the IR beamline available at DAΦNE. The array has been aligned at the focus spot of the optical system and placed after the last toroidal mirror of the beamline. The SINBAD optical system de-magnifies the source image by a factor 2.3 and at mid-IR wavelengths its size is roughly ~ 2.0 x 1.5 mm$^2$ at the final focus.

The array has been placed in the vertical plane in front to the IR spot. We focused the beam spot on the array using the last optical element of the beamline, fully illuminating the four bonded pixels of the array in the vertical, direction optimizing signals until to reach with a minimum current of the beam a S/N ≥ 10. However, although the total active area of the array is ~2300x100 μm$^2$, at present our set up can monitor only a limited portion of the spot, ~500x100 μm$^2$.

Signals have been collected by four input channels of a WaveRunner 6100A of LeCroy scope with a bandwidth of 1 GHz and at 10 Gsample/s. Acquisitions of the four channels have been carried out simultaneously using a standard GPIB-port with a software package under the LabView platform that allows the simultaneous acquisition and saving data. In the future, to collect and store simultaneously signals from all channels we plan to use a fast digital electronics, FPGA based, now under test.

## RESULTS

Before the shut down of the DAΦNE machine, in the summer of 2009, different measurements with the e$^-$ beam have been performed with this IR array. Experiments have been carried out during different runs of DAΦNE acquiring data in more than one consecutive bunch re-filling of the e$^-$ beam.

An example of data acquisition obtained with these four pixels of the array is showed in Fig. 1. In this figure each signal shows the IR emission of all 100 bunches circulating and also the bunch gap is clearly resolved. The signal of the first pixel is reported with a blue line, the second with a red line, the third with a black line and finally the fourth pixel with a green line. To reduce the noise level and to obtain a higher S/N the acquisition has been performed averaging 32 sweeps of each waveform.

In Fig 2 a magnified view of the previous plots is given showing the single bunch signals with part of the gap in the time structure. Signals in Fig.2 show that each pixel of the device, each one with a response time of about ~1-1.2 ns, clearly resolve the emission of each bunch and the separation among them of 2.7 ns. However the response time of the device is not fast enough to allow the IR signal to reach the offset level observed at the beginning of the first bunch.

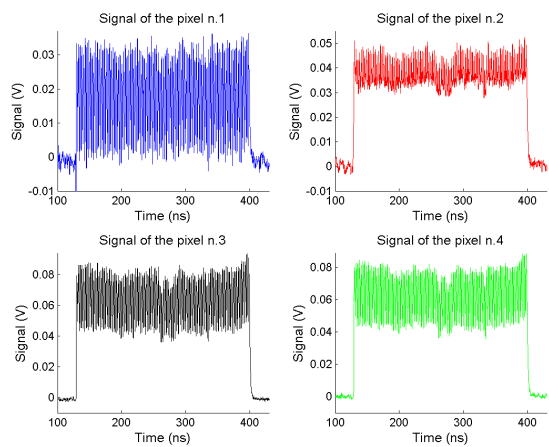

Figure 1: Comparison of the simultaneous acquisition of from the pixels. Data collected at SINBAD allow monitoring a portion of the electron bunch IR emission.

Fig. 2 shows also that signals from these pixels have different amplitudes and that with such a device is possible to monitor in real time, bunch-by-bunch the shape of the spot.

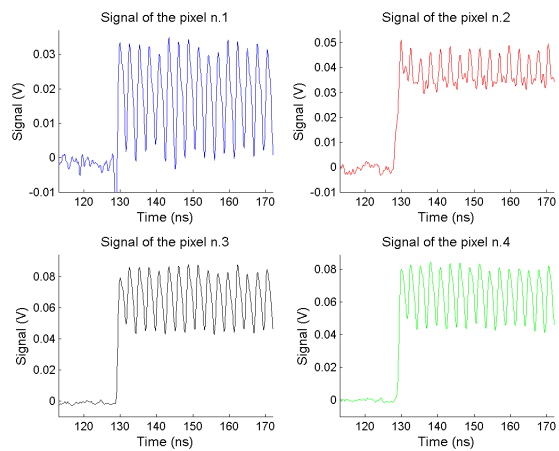

Figure 2: Magnified view of the signals in Fig. 1. In each panels we resolve the IR emission of the first bunches of the beam and the gap structure.

In Fig. 3 is showed the behaviour of signals of the first bunch vs. time, as seen by the four pixels. In the upper panel in Fig. 3 we compare signals of the first electron bunch of the train as collected by the four pixels during two consecutive runs of DAΦNE. Below, in the second panel we show differences between the signals of the first bunch collected by the different pixels: the first minus the third, the first minus the forth and the second minus the third. In the third panel of the same figure the current of the first bunch is also showed for comparison. We have to underline here that the behaviour of the signal of all bunches of the train can be plotted and analysed.

Curves in the upper panel in Fig.3 clearly points out amplitude differences among signals measured by the different pixels of the array during the data acquisition. Also at first glance the comparison among IR signals and bunch current (in the third panel) shows a high correlation among the observed current behaviour and IR emissions. In the second panel the amplitude differences among pixels also show that small variations in the IR signals associated to portions of the IR spot can be monitored. In particular we like to underline that these small variations in the amplitudes during the run can allow correlations among changes in the IR emissions of the source with displacements in the position of the spot with respect to the illuminated pixels.

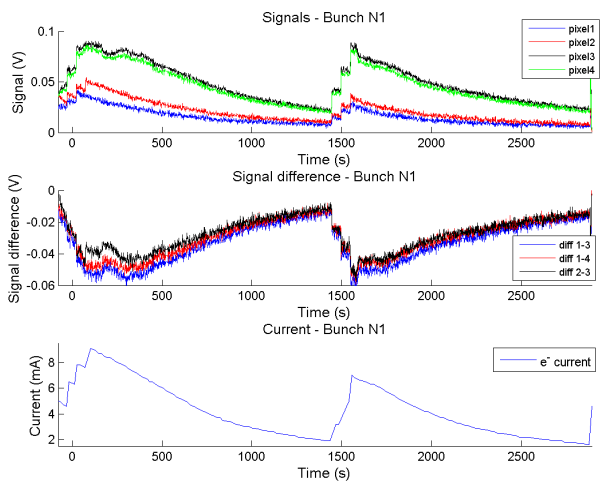

Figure 3: Comparison among pixel signals of the array of the first bunch (upper panel); comparison among differences between pixels of the array (middle panel); the bunch current (lower panel) collected during two consecutive runs at DAΦNE.

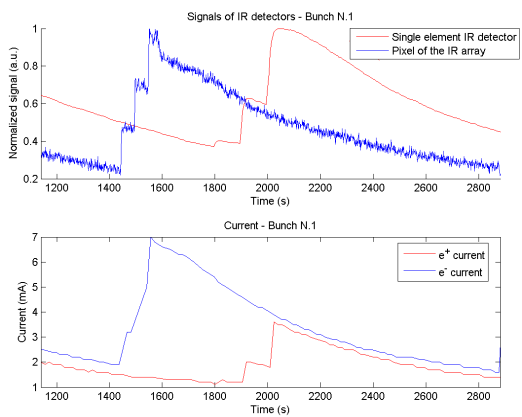

Figure 4: Comparison among the signals collected by one pixel from an e⁻ beam (blue) and a single element detector monitoring at the same time the e⁺ beam (red) during a DAΦNE run (upper panel) compared with the bunch current behaviours (lower panel).

Such device has been designed for the final installation in the framework of the 3+L experiment, an experiment dedicated to the beam diagnostic of the $e^+$ ring at DAΦNE. A new IR port and an associated experimental set up designed to perform bunch-by-bunch diagnostics of the $e^+$ beam of the DAΦNE collider, has been recently installed and commissioned [3]. As an example, signals of a single bunch collected by different IR fast detectors are showed in Fig. 4. Data of the IR $e^-$ emission (blue) are compared with the signal detected by a fast single element IR detector, installed at the focus of the 3+L optical systems collecting the $e^+$ emission (red). In lower panel in Fig.4 both $e^-$ and $e^+$ bunch currents (blue and red lines, respectively) are also showed. Looking at Fig. 4 it is also clear that smaller variations of the signal during runs can be monitored by a single pixel of the array with a smaller area (50x50 $\mu m^2$) with respect to the single element IR detector actually used (~1 $mm^2$).

## CONCLUSIONS

First measurements performed monitoring signals of four pixels of a fast IR array demonstrate that monitoring a source at IR wavelengths in the sub-ns time scale is possible. A simple image of the spot can be also obtained. Moreover, small amplitude variations of the pixel signal can be correlated to possible changes of the characteristics of the source and/or to beam displacements in the sub-ns scale. To fully reconstruct the dimensions and the shape of the spot, acquisitions of a large number of pixels is necessary. To this purpose an improvement of the method of the analysis of the large amount of data collected with the array is still in progress. Improvements based on a dedicated electronics and a dedicated digital electronics, based on a powerful FPGA chip, is in progress. The system will allow to collect and store in real time signals from all the 64 channels of the IR array. Dedicated experimental run are foreseen at DAΦNE after the ongoing upgrade of the storage ring complex and also at other particle accelerators, such as the Hefei Synchrotron Source in China [4].

Our data show that, in principle, new array devices at different wavelengths with faster response times and better performances, in terms of signal to noise ratio may be designed, assembled and characterized. These optimized devices should be reliable used for real time diagnostic of the transverse dimensions of particle beams. In particular such new diagnostic, thanks to the possibility to correlate in a real time the position of the bunch of each packet in the train, appears suitable to collect bunch-by-bunch and turn-by-turn data and used to extract direct and precious information regarding beam instabilities.